\documentclass{elsart}

\usepackage{natbib}

\usepackage{graphicx}
\usepackage{amssymb}

\begin{document}

\begin{frontmatter}


%\title{Non-collinear magnetism in iron at high pressure}

\title{Non-collinear magnetism in iron\\
at high pressures}


\author{R. E. Cohen}
\ead{cohen@gl.ciw.edu}
\address{Geophysical Laboratory, Carnegie Institution of Washington\\
5251 Broad Branch Rd., N.W., Washington, D.C. 20015}
\begin{abstract}

Using a first principles based, magnetic tight-binding total energy
model, the magnetization energy and moments are computed for
various ordered spin configurations in the high pressure polymorphs of iron (fcc, or $\gamma$-Fe, and hcp, or $\epsilon$-Fe), as %%@
well ferromagnetic bcc iron ($\alpha$-Fe). For hcp, a non-collinear, antiferromagnetic, spin configuration that minimizes %%@
unfavorable ferromagnetic nearest neighbor ordering is the lowest energy state and is more stable than non-magnetic $\epsilon$ iron %%@
up to about 75 GPa.  Accounting for non-collinear magnetism yields better agreement with the experimental equation of state, in %%@
contrast to the non-magnetic equation of state, which is in poor agreement with experiment below 50 GPa.

\end{abstract}


\end{frontmatter}

% main text
\section{Introduction}
Magnetism is known to be important in the phase stability, structure and elastic properties of iron. For example, $\alpha$-Fe, the %%@
ground state at ambient conditions, would be mechanically unstable if it were not magnetic. Even above the Curie temperature, T$_c$ %%@
there are local magnetic moments in $\alpha$-Fe. Face-centered cubic iron (fcc or $\gamma$-Fe) has incommensurate magnetic %%@
correlations which change rapidly with volume, and give rise to the anti-Invar effect (large thermal expansivity) \citep{fcc}. The %%@
magnetic behavior of hcp iron is important for high pressure materials
research, and for interpreting high pressure experiments aimed at understanding the Earth's inner core. Hcp iron is not quenchable %%@
to zero pressure, so magnetic studies must be made {\it in situ} at high pressures. $\epsilon$-Fe was long thought to be %%@
non-magnetic, due to several Mossbauer experiments that showed no hyperfine splitting in $\epsilon$-Fe, even down to helium %%@
temperatures. However, self-consistent first-principles computations show a magnetic ground stare for $\epsilon$-Fe which is stable %%@
up to about 50 GPa \citep{gerd}. Here non-collinear magnetism in $\epsilon$-Fe is explored using a magnetic tight-binding model %%@
fitted to first-principles calculations. 

When magnetic moments are collinear, electrons can be considered to be ``spin-up" or ``spin-down" in a global sense, i.e. there is %%@
a global magnetic quantization direction. This means that one can solve for the spin-up and spin-down electrons separately, and %%@
then combine the results to compute the total charge density. When moments are oblique to each other, the spin state is said to be %%@
``non-collinear." In that case the problem does not factorize, and one must diagonalize a Hamiltonian of twice the order of the %%@
collinear case. A system can be non-collinear either in response to chemical or thermal disorder, or in order to minimize %%@
frustration. Both hcp and fcc lattices are frustrated with respect to antiferromagnetism, in that one cannot order these lattices %%@
antiferromagnetically so that all neighbors have opposite spins. This is known to lead to spin-waves and non-collinear magnetism in %%@
fcc-Fe. Also, a common example of non-collinear spins occurs on heating bcc-Fe above the Curie point, where the spins disorder %%@
dynamically. A comprehensive review of the theory of non-collinear magnetism is given in \citet{sandratskii}.

\section{Method}

To study non-collinear magnetism in Fe, a first-principles based non-magnetic tight-binding model %%@
\citep{ronnonmag1,ronnonmag2,ronnonmag3} is combined with a model for magnetism \citep{pickett,sonali}. The Hamiltonian (H) of the %%@
system
is given by,
\begin{equation}
\label{eq:hamiltonian}
H = H_0 + R^{+}H_{collinear}R,
\end{equation}
where $H_0$ is the (doubled) non-magnetic Hamiltonian, $H_{collinear}$ is the collinear magnetic Hamiltonian and $R$ is the %%@
rotation matrix for spin directions, as described below. The overlap matrix $S$ is unchanged from the non-magnetic non-orthogonal %%@
tight-binding model \citep{ronnonmag1,ronnonmag2,ronnonmag3}. $H_0$ was fit to
eigenvalues and total energies from an extensive set of non-magnetic Linearized Augmented Plane Wave (LAPW) results within the %%@
Generalized Gradient Approximation (GGA) \citep{PBE} in the bcc, fcc, and hcp
structures as functions of pressure and strain, and has been extensively tested \citep{ronnonmag1,ronnonmag2,ronnonmag3}.

The magnetic Hamiltonian is twice the size of the original non-magnetic Hamiltonian, with the upper left block being spin-up, the %%@
lower right block is spin down, and the off-diagonal blocks are from coupling between up and down spins that give rise to %%@
non-collinear magnetic solutions. $H_0$ is doubled from the original non-magnetic Hamiltonian, with identical diagonal blocks, and %%@
zeros on the off-diagonal blocks. The only non-zero elements in the collinear magnetic Hamiltonian H$_{collinear}$ are the diagonal %%@
onsite $d$ elements. The latter are given by $-I m_i/2$ for the spin-up block, and $+I m_i/2$ for the spin-down block, where $m_i$ %%@
is the magnetic moment of the atom $i$, and $I$ is the Stoner parameter, which controls the strength of the exchange splitting %%@
\citep{pickett}. $R$ is the rotation matrix which depends upon the atomic spin direction with respect to a global reference frame %%@
\citep{uhl}. The diagonal of $H_{mag}$ is spin-up and spin-down, and the off-diagonal bands are the coupling between up and down, %%@
which arise from the rotation operation. Similar models have been used previously for incorporating magnetism within a %%@
tight-binding approach \citep{you,pickett,freyss,mehl}, but the magnetic behavior of
$\epsilon$-Fe has not been addressed, nor has a first-principles tight-binding model previously been developed that includes %%@
constraining fields.

The full spin density matrix is computed, and the constraining fields are computed self-consistently so that spin directions as %%@
well as spin moments are self-consistent. The spin-density matrix is given by
\begin{equation}
\rho_{\sigma\sigma^\prime}=\sum_{ij} \psi_{i\sigma} S_{ij} \psi_{j\sigma^\prime}
\end{equation}
where $S_{ij}$ is the overlap matrix and $\psi_{i\sigma}$ are the eigenvectors for atom orbital $i$ and spin $\sigma$ in the local %%@
coordinate system on each atom. In the local coordinate system the spin on each atom is diagonal and can be represented as up or %%@
down. The spin-1/2 rotation matrix elements of $R$ that rotates from the local coordinate system to the global system with oblique %%@
spins is given by
\begin{equation}
R = \left( \begin{array}{cc} 
R_{\uparrow\uparrow} & R_{\uparrow\downarrow} \\
R_{\downarrow\uparrow} & R_{\downarrow\downarrow} \\
\end{array} \right)
=\left( \begin{array}{cc}
\cos \frac{\theta}{2} e^{i\frac{\phi}{2}} & \sin \frac{\theta}{2} e^{-i\frac{\phi}{2}}  \\
-\sin \frac{\theta}{2} e^{i\frac{\phi}{2}} & \cos \frac{\theta}{2} e^{-i\frac{\phi}{2}}  \\
\end{array}
\right)
\end{equation}   
where $\theta$ is the polar angle and $\phi$ is the azimuthal angle. The matrix elements given here are used to build the full %%@
rotation matrix; the only non-zero elements are diagonal in atom orbital indices. Without spin-orbit coupling, there is no coupling %%@
to absolute directions relative to the crystalline lattice, and only the relative angles between spins are important. Spin-orbit %%@
interactions are not included here; their energetic importance is insignificant compared with the exchange energy changes for iron.

The magnetic moments are given by the spin density on each atom summed over the orbitals. The expressions are
\begin{eqnarray}
m & = & \sqrt{(\rho_{\uparrow\uparrow}-\rho_{\downarrow\downarrow})^2 +4 \rho_{\uparrow\downarrow}\rho^*_{\downarrow\uparrow}} \\
\theta & = & \arccos \left( \frac{\rho_{\uparrow\uparrow}-\rho_{\downarrow\downarrow}}{m} \right) \\
\phi & = & \Im \log \rho_{\uparrow\downarrow} .
\end{eqnarray}

The magnetic moments are determined self-consistently. One starts with an initial guess, which gives the exchange splittings, $\pm %%@
I m_i/2$, in the Hamiltonian, find $H$, and diagonalize the generalized eigenproblem. States are occupied up to the Fermi level, %%@
and the output moment on each atom is found from the weighted eigenvectors and the expressions above. The process is repeated until %%@
self-consistency is obtained. 

In general the output moment direction, as well as magnitude, will differ from the input.  In this way a self-consistent process %%@
can be used to find the moment magnitudes and directions that are locally stable. In some cases one wants to compute the energy of %%@
a given magnetic structure. In order to guarantee that the output magnetic structure is the desired structure, it is necessary to %%@
apply magnetic fields that force the output moments to form the desired structure.  Such a procedure has been implemented using a %%@
self-consistent procedure with the option to constrain the moments in magnitude and direction (constraining fields contain %%@
longitudinal components), or to allow the magnitude to adjust self-consistently, but force the direction to remain as desired %%@
(constraining fields are transverse). The procedure used is similar to, but different from, the procedure outlined in %%@
\citet{ujfalussy}.

The total energy is given by 
\begin{equation}
\label{eq:totale}
E_{total}=E_b+I \sum m_i^2/4 + \sum \vec{b}_i \cdot \vec{m}_i 
\end{equation}
where $E_b$ is the band energy (sum of the eigenvalues). There are no additional potential terms; the total energy is obtained from %%@
the band structure and moments. The second term in Eq.~\ref{eq:totale} corrects for double counting of the exchange interaction %%@
\citep{pickett}, and the third term corrects for double counting of the interaction with the applied staggered fields $\vec{b}_i$.

The model differs from the conventional Stoner model \citep{stoner}, in which a ferromagnetic instability is predicted by the %%@
inequality $I N(0) > 1$, where $N(0)$ is the non-magnetic density of states at the Fermi level, and the extended Stoner model %%@
\citep{krasko}, in which $N$ is replaced by the effective density of states $\tilde{N}(M)=M/\delta \epsilon$, where $\delta %%@
\epsilon$ is the exchange splitting, in that our model allows for different hybridization depending on magnetic state, accounting %%@
thereby to the actual magnetic structure, i.e. whether the system will be ferromagnetic, antiferromagnetic, or non-collinear.

\section{Results and Discussion}
\subsection{bcc}
\begin{figure}[tbp]
\centering
\includegraphics[width=4in]{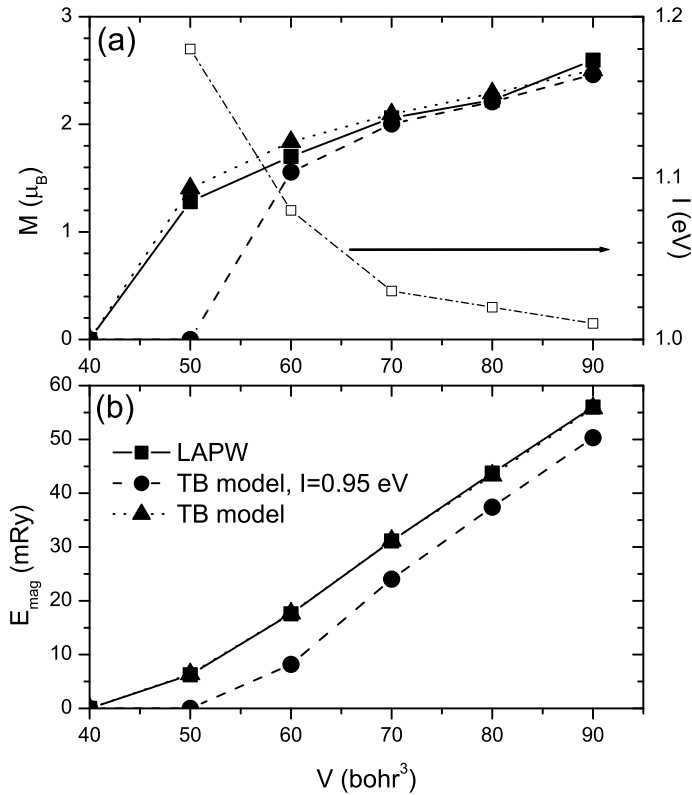}
\caption{\label{fig1}
Magnetization (a) and magnetization energy (b) for bcc iron as functions of volume. The self-consistent LAPW results are from %%@
\citet{bcc}. The tight-binding (TB) model results are shown for constant $I$=0.95 eV, and for varying $I$, with $I$ chosen to best %%@
fit the LAPW magnetization energies. The best-fit $I$ is also shown in (a).}
\end{figure}

The model is in good agreement with previous
self-consistent calculations for $\alpha$-Fe, especially when a volume dependent $I$ is used \citep{sonali}.  The lowest energy
state for $\alpha$- Fe for the model at ambient and at higher pressures is
ferromagnetic, in agreement with the first principles LAPW self-consistent calculations \citep{bcc}.
Moreover, even with a single value of volume independent $I$ the model
calculation describes well the pressure dependence of the magnetic moment
and the magnetization energy. As pressure is increased
the magnetization energy decreases smoothly (Fig.~\ref{fig1}). The value of $I$ needed for
quantitative agreement is almost constant (around 1 eV) for larger
atomic volumes or lower pressures but increases for higher
pressures. This is consistent with the increase in the exchange interaction at higher pressures noted previously \citep{asada}. %%@
Even though the exchange interaction
increases with pressure, the non-magnetic density of states near
the Fermi energy decreases, giving rise to a net
reduction in the magnetic moment and magnetization energy, and finally
the loss of magnetism at around atomic volume of 40 bohr$^3$.

The Stoner parameter $I$ was adjusted at each volume to give agreement with self-consistent full-potential Linearized Augmented %%@
Plane Wave (LAPW) magnetization energies \citep{gerd} for each volume for bcc. $I$ is expected to increase as the electronic %%@
density increases with decreasing atomic volume, similar to what has been seen
before for $\gamma$-Fe \citep{krasko} and $\alpha$-Fe (Fig.~\ref{fig1}). Fitting the results to a polynomial valid between $V=$50 %%@
bohr$^3$ and 90 bohr$^3$ gives \(I \textstyle{(eV)}= 3.4126 - 0.08583 V + 0.00103 V^2 - 4.1666 10^{-6} V^3\) .

\subsection{fcc}
\begin{figure}[tbp]
\centering
\includegraphics[width=5.5in]{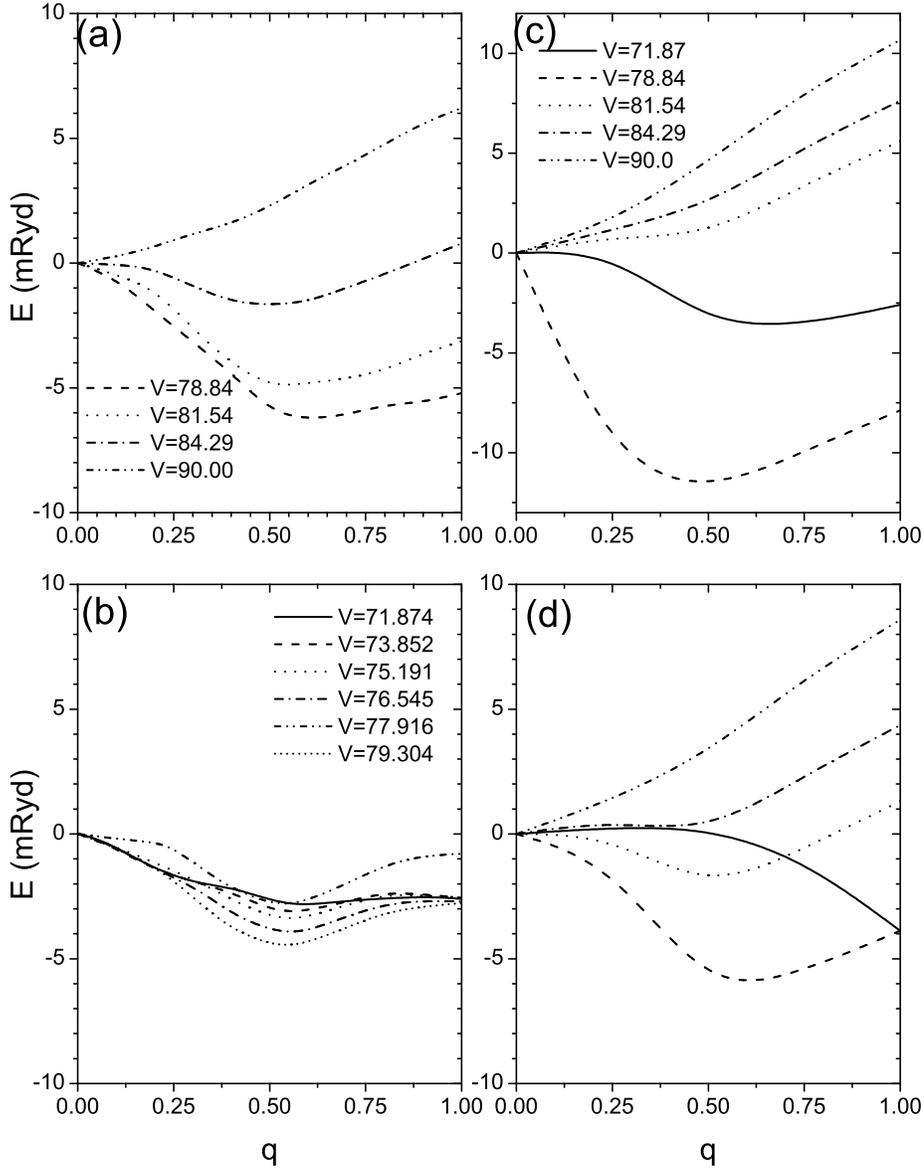}
\caption{\label{fig:fcce}
Variation of magnetization energy versus wavevector for $\gamma-Fe$ for (0,0,q), with $q$ in units of $2\pi/a$ , where $a$ is the %%@
lattice constant).  Wavevectors $q=0$ and $q=1$
correspond to ferromagnetic and antiferromagnetic structures, respectively, $q=0.5$
is a non-collinear spin state with relative angle between the
neighboring spins of $\pi/2$. (a) LMTO results from \citet{fcc}. (b) LAPW results from \citet{sjostedt}. (c) Results from this %%@
study using the best-fit value of $I$ (1.017, 1.006, 1.003, 1.000, and` 0.993 eV for the volumes shown). (d) Results from this %%@
study using $I=$0.94 eV. The tight-binding results show the correct behavior of ferromagnetic at large volumes and antiferomagnetic %%@
at small volumes (high pressures) with a non-collinear transition region. The results of \citet{sjostedt} seem more different than %%@
our results and those of \citet{fcc} than can be explained by the atomic moment approximation. Note that the line dashes and %%@
volumes correspond for (a), (c) and (d) but not for (b).} 
\end{figure}

\begin{figure}[tbp]
\centering
\includegraphics[width=5.5in]{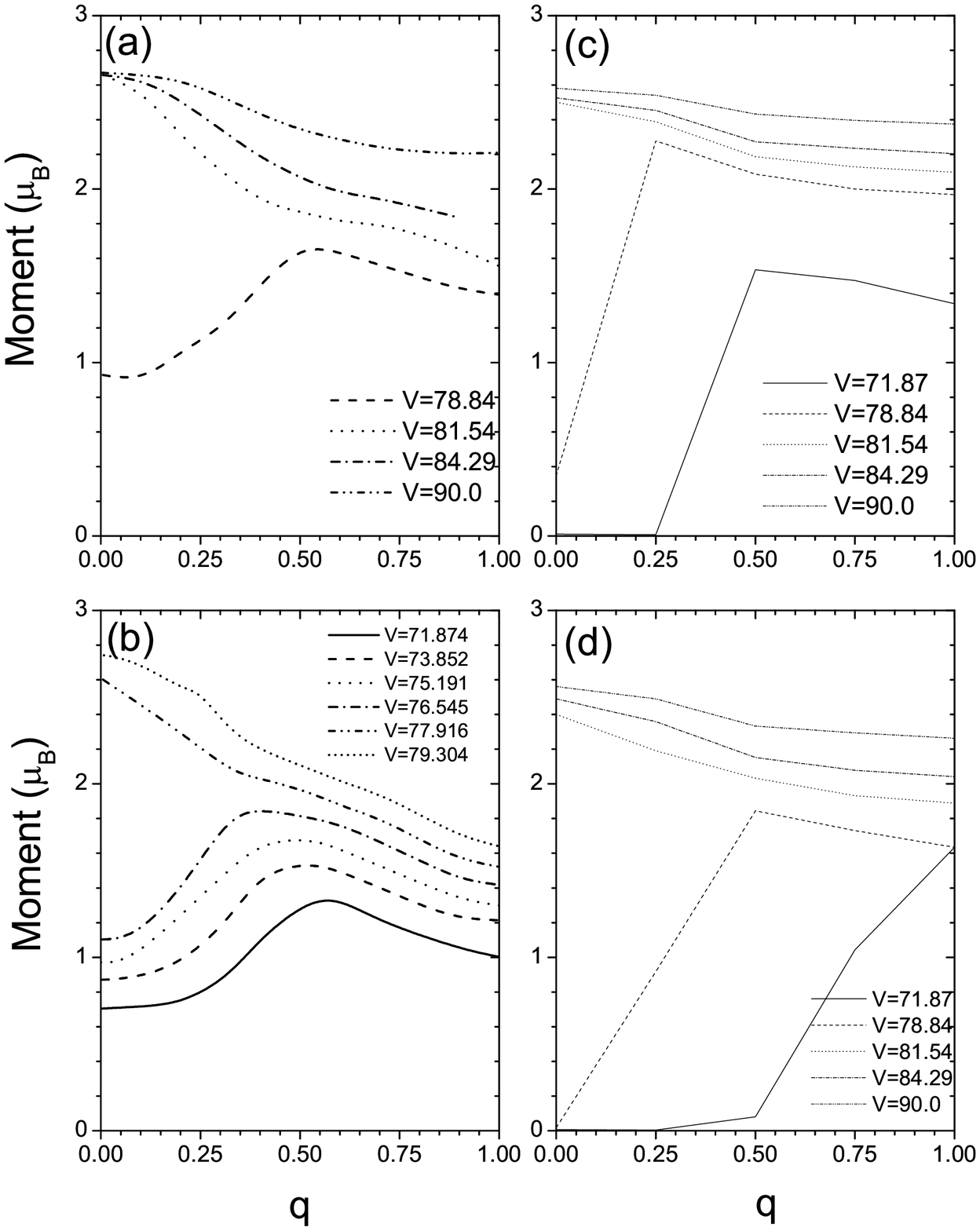}
\caption{\label{fig:fccmom}
Variation of moments versus wavevector for $\gamma-Fe$ for (0,0,q). (a) LMTO results from \citet{fcc}. (b) LAPW results from %%@
\citet{sjostedt}. (c) Results from this study using the best-fit value of $I$ (1.017, 1.006, 1.003, 1.000, and` 0.993 eV for the %%@
volumes shown). (d) Results from this study using $I=$0.94 eV. Note that the line dashes and volumes correspond for (a), (c) and %%@
(d) but not for (b). A high-spin low-spin transition is evident in all cases for ferromagnetic and low-$q$ spin-waves with %%@
increasing pressure. The tight-binding model apparently tends to give too small a moment in the low spin regime, but is in %%@
generally good agreement.}
\end{figure}

$\gamma$-Fe shows a great richness in non-collinear magnetic structures as volume is varied, and here it is not explored in detail;  %%@
much study has been done, as reviewed in \citet{sandratskii}. The energies and moments for fcc-Fe with spiral spins along (001) and %%@
$\theta=\pi/2$ were determined using eight-atom supercells, allowing calculations for $(00q)$ with $q=0,\pi/4,\pi/2,3\pi/4$,and %%@
$\pi$. A k-point mesh of 12$\times$12$\times$4 was used, giving 72 k-points in the irreducible wedge for tetragonal symmetry. For %%@
$\gamma$-Fe, the complex behavior of magnetic moments and the
magnetization energy with increasing pressures and varying spiral spin
density wave states is qualitatively reproduced by the model (Figs.~\ref{fig:fcce} and \ref{fig:fccmom}) compared with %%@
self-consistent calculations \citep{fcc,uhl,sjostedt}. Quantitatively the results are sensitive to the value of $I$, and to get %%@
good agreement with self-consistent calculations it seems a smaller I, about 0.94 eV, is required than derived from bcc-Fe %%@
(0.99-1.02 eV for the volumes considered for fcc). In any case there is some variation in self-consistent results for %%@
$\epsilon$-Fe, due to extreme sensitivity of the magnetic structure to basis set, k-point sampling, etc., due to the small energy %%@
scale. Furthermore, the magnetic ground state in $\gamma$-Fe is sensitive to the atomic moment approximation \citep{sjostedt} in %%@
the tight-binding model; that is the moment is really a field that varies with position in the crystal, and is not constant on each %%@
atom. Within the atomic moment approximation our results are reasonably consistent with the self-consistent results. More detailed %%@
comparisons with the comprehensive non-collinear LAPW calculations of \citet{sjostedt} including k-point convergence tests are %%@
called for, but have not yet been done. In  \citet{sjostedt} 4000 k-points were used in the full Brillouin zone, compared to our %%@
4608, which seems comparable. It is more difficult to compare with \citet{fcc} since a real space multiple scattering approach and %%@
a muffin-tin potential approximation was used.

\subsection{hcp}
\begin{figure}
\centering
\includegraphics[width=2.5in]{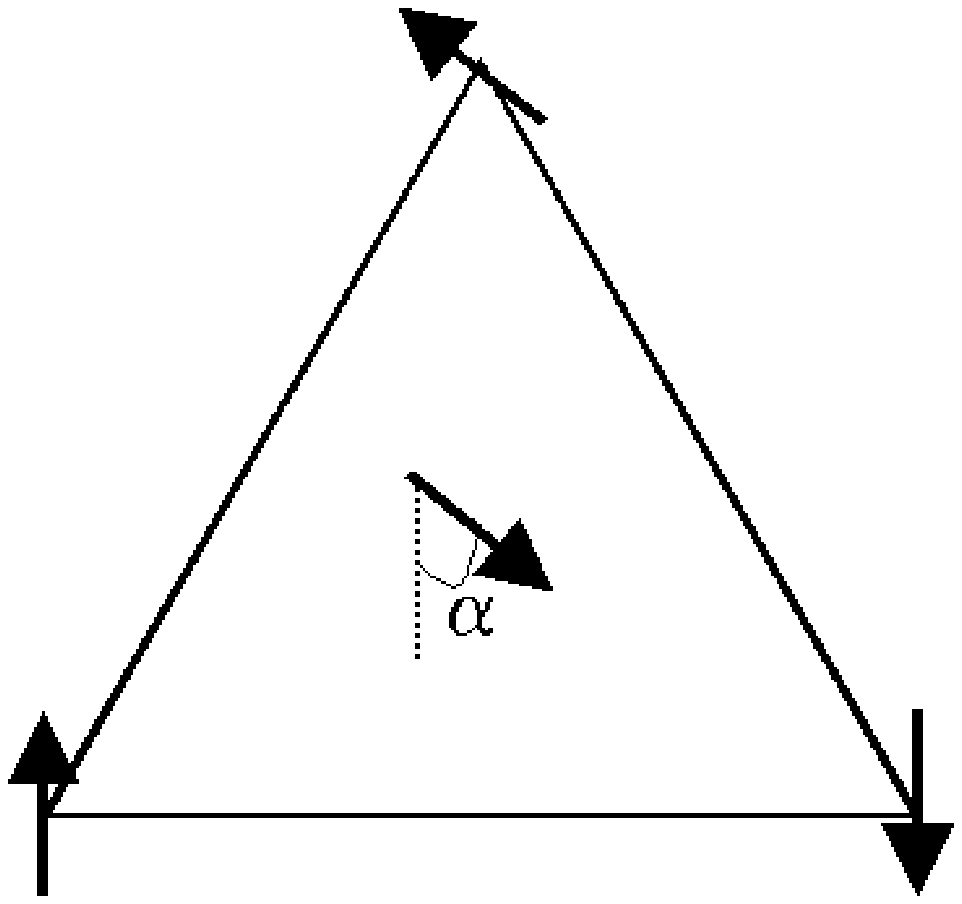}
\caption{\label{fig:angle}
The motif for an hcp lattice with antiferromagnetic interactions. One pair of antiferromagnetic iron atoms is at an angle $\alpha$ %%@
to another pair. In the Heisenberg model with near-neighbor interactions, the total energy ins independent of $\alpha$. The %%@
collinear afmII structure is represented by $\alpha=0$.} 
\end{figure}

Previous collinear first principles calculations show stability of an
antiferromagnetic state, afmII, in hcp iron \citep{gerd}. Moreover,
the computed equation of state of afmII greatly improves agreement
with the experimental equation of state. An hcp lattice is frustrated for antiferromagnetism; it is not possible to have perfect %%@
antiferromagnetic order on it. In the afmII structure each atom has eight antiferromagnetically oriented and four ferromagnetically %%@
oriented neighbors, maximizing the antiferromagnetic interactions. In the case of a nearest-neighbor (n.n.) Heisenberg model, which %%@
has energy $E = J_1 \sum_{n.n.} \vec{m}_i \cdot \vec{m}_j $, the energy is independent of the angle between the moments of one %%@
antiferromagnetic pair and another (see Fig.\ref{fig:angle}). If non-neigherest neighbor interactions are important, or if the %%@
Heisenberg model does not completely describe the energetics, the energy might be further lowered if the antiferromagnetic pairs %%@
are oblique or perpendicular to each other (i.e. $\alpha \neq 0$). 

\begin{figure}
\centering
\includegraphics[width=3.5in]{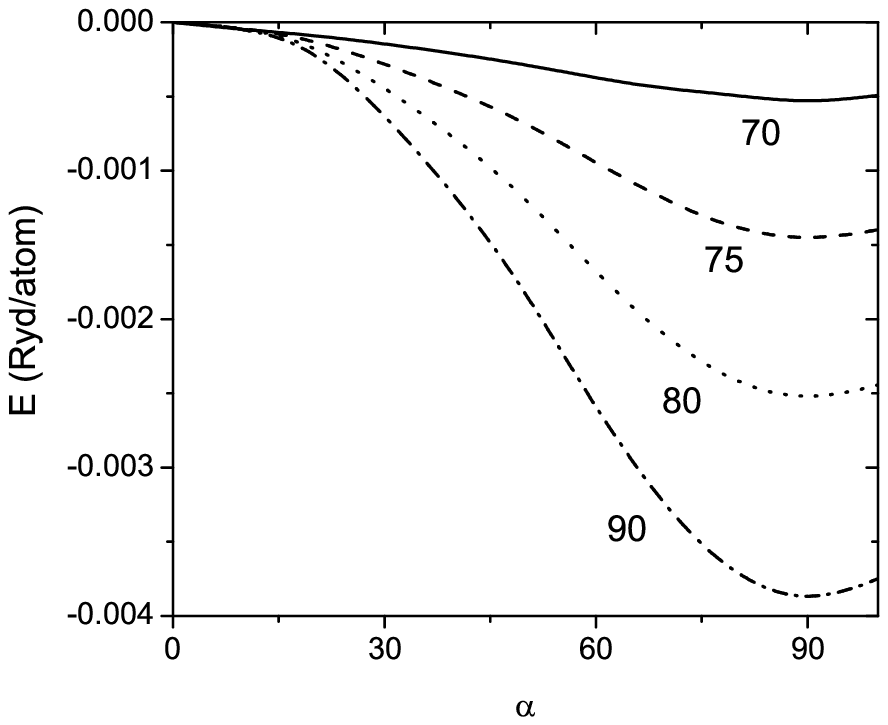}
\caption{\label{fig:rotate}
The motif for an hcp lattice with antiferromagnetic interactions. One pair of antiferromagnetic iron atoms is at an angle $\alpha$ %%@
to another pair. In the Heisenberg model with near-neighbor interactions, the total energy ins independent of $\alpha$. The %%@
collinear afmII structure is represented by $\alpha=0$.} 
\end{figure}

The collinear afmII structure has 4 atoms per unit cell in space group Pmma \citep{gerd}. In order to tile the lattice with the %%@
pattern shown in Fig.\ref{fig:angle} a cell with 8 atoms is obtained with space group  A value of 1.6 was used for $c/a$ for all of %%@
the hcp based calculations here, which is close to the ground state value for the volumes studied here. Tests showed that $c/a$ %%@
does not vary significantly (<.005) with magnetic state.  Figure \ref{fig:rotate} shows the energy versus angle $\alpha$ for %%@
different volumes. In all cases the energy decreases when $\alpha$ is varied from zero, indicating that the ground state is %%@
non-collinear. The most stable state is with $\alpha$=90$^\circ$, which minimizes the local ferromagnetic interactions. Table %%@
\ref{table:e} shows the results of the TB model for hcp Fe with ferromagnetic, admII, and the non-collinear structure with %%@
$\alpha$=90$^\circ$ (ncl). 

\begin{table}[tbp]
\caption{
\label{table:e}
Magnetization energy (mRy/atom) for different atomic volumes V (bohr$^3$). 
The value of Stoner parameter $I$ (eV) for the different volumes 
is given in the first column.  The magnetic moments are given in parenthesis. The non-collinear (ncl) structure is for %%@
$\alpha$=90$^\circ$} 
\begin{tabular}{ccccc}
V (au)&I (eV)&Ferro&afmII&ncl\\
\hline
90& 1.01& 27.1 (2.6) & 17.7 (2.28) & 21.6 (2.38) \\
80& 1.05& 7.2 (2.6) & 7.1 (1.79)& 9.7 (1.99)\\
75& 1.1& 0 (0) & 3.1 (1.47)& 4.6 (1.67)\\
70& 1.14&0 (0) & 0.55 (0.93)& 1.1 (1.13) \\
65& 1.19&0 (0) & 0.08 (0.33) & 0.02 (0.19)
\end{tabular}
\end{table}

\begin{figure}[tbp]
\centering
\includegraphics[width=3.5in]{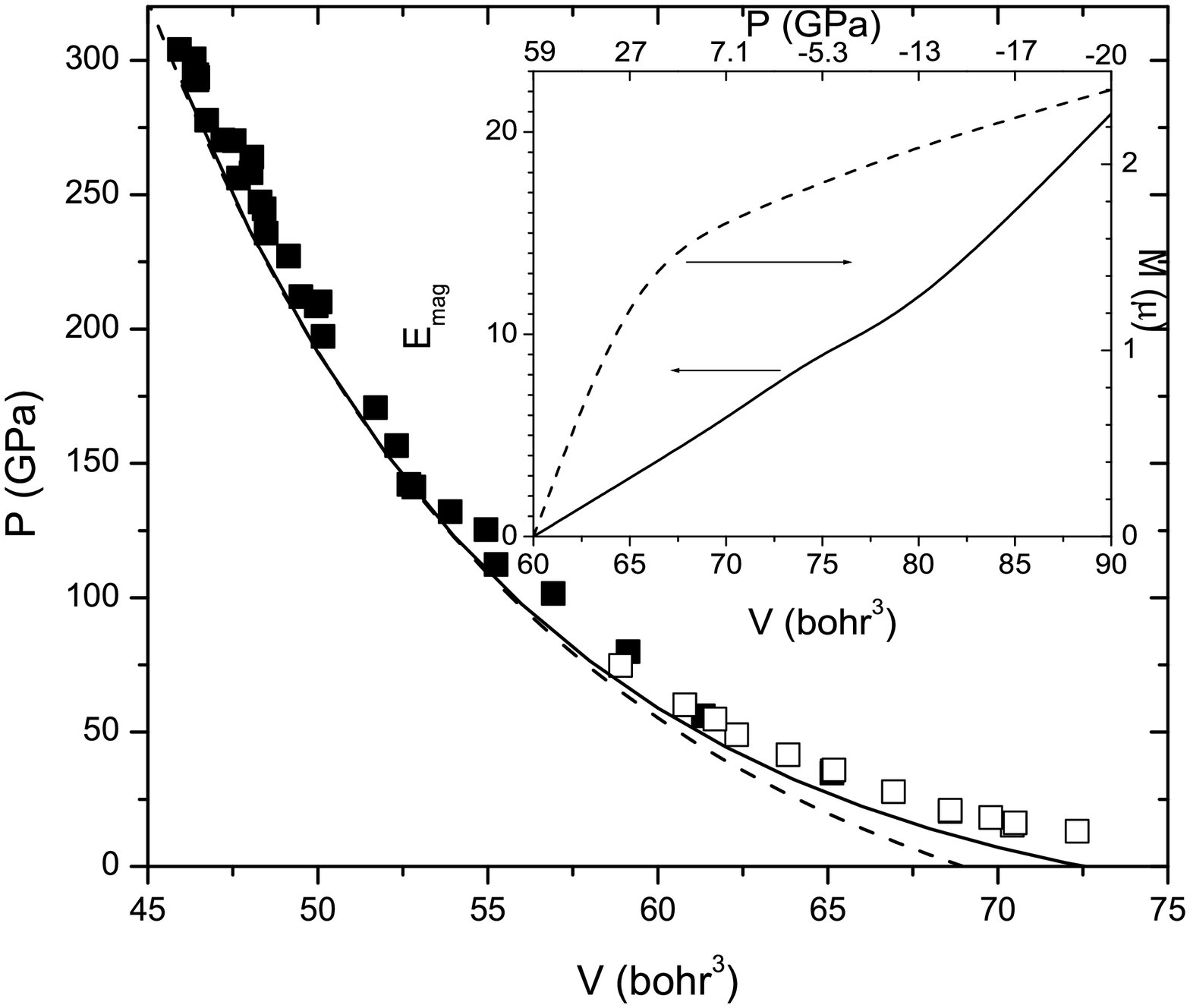}
\caption{\label{fig5}
Equation of state of $\epsilon$-Fe. Symbols are experiments from \citet{jephcoat} (open) and \citet{mao} (solid). The solid and the %%@
dashed curves are our magnetic and non-collinear NC 1 magnetic theoretical tight-binding results.} Inset shows the magnetization %%@
energy per atom and moments in NC 1 versus volume. The upper non-linear scale shows the corresponding theoretical pressures.
\end{figure}

The ground state non-collinear total energies of $\epsilon$ -e
were fit to a Vinet equation of state \citep{vinet} as a function of volume. Table \ref{table:eos} shows the results both using %%@
$K^\prime$=4 and allowing $K^\prime$ to vary. For non-magnetic Fe, $K^\prime$=4 is not a bad approximation, but when magnetism is %%@
included $K^\prime>$6. The results show that the TB model is in god agreement with the self-consistent LAPW computations of %%@
\citet{gerd}. Including magnetism lowers the bulk modulus, and including non-collinear magnetism lowers it further. The resulting %%@
equation of state fo ncl is in very good agreement with the
experiments (Fig.~\ref{fig5},Table \ref{table:eos}) \citep{jephcoat,mao}.  When magnetization is not included the
disagreement between the experiment and theory is 75$\%$ and 9$\%$ for
the bulk modulus and the equilibrium volume respectively. Including
the afmII structure as the ground state of hcp reduces the
disagreement to 25$\%$ and 5$\%$ for bulk modulus and equilibrium
volume \citep{gerd}. When the non-collinear spin state energies were
used in the equation of state the bulk modulus was within 10$\%$ of
the experiment.

\begin{table}[tbp]
\caption{Comparison of experimental and theoretical values of equilibrium 
volume ($V_0$) and bulk modulus ($K_0$) for $\epsilon$-Fe.} 
\label{table:eos}
\begin{tabular}{llll}
Fe (GGA)&$V_0$ (bohr$^3$)&$K_0$(GPa)& K$^\prime$\\
\hline
Expt\citep{mao}&75.4&165&\\
Non-Magnetic (GGA) \citep{gerd}&69.0&292&4.4\\
Non-Magnetic (this study)&69.1&300&4.0\\
Non-Magnetic (this study)&68.8&297&4.6\\
Collinear (afmII) \citep{gerd}&71.2&209&5.2\\
Collinear (afmII) (this study)&71.3&240&4.0\\
Collinear (afmII) (this study)&70.1&214&6.3\\
Non-Collinear (this study)&72.0&227&4.0\\
Non-Collinear (this study)&70.7&195&6.5\\
\end{tabular}
\end{table}

In spite of theoretical evidence for magnetism in $\epsilon$-Fe, and the great
improvement in the equation of state when magnetism is included, the experimental situation is unclear. M\"{o}ssbauer spectroscopy %%@
shows no evidence of magnetism in hcp Fe \citep{taylor,jeanloz}. X-ray absorption spectroscopy also has been interpreted to show %%@
lack of magnetism  \citep{rueff}, but due to the absence of an absolute calibration,
and sensitivity of the spectrum to changes in the density of states, the experiments show only
that magnetic moments are lower in $\epsilon$-Fe than $\alpha$-Fe, a
result consistent with theory. On the other hand, there is some
independent evidence of magnetism from Raman spectroscopy, which shows
two peaks \citep{merkel}, instead of the one expected in the hcp
structure. The frequencies and splitting of these peaks is predicted
well from first-principles calculations for magnetic ordered afmII
$\epsilon$-Fe \citep{gerd2}, and the absence of observed splitting in M\"{o}ssbauer will be explained in a subsequent paper.

Thanks to B.\ Fultz, A.\ Goncharov, R.\ Hemley, A.\ Jephcoat, I.\ Mazin, H.\ Olijnyk, W.\ Pickett, and G.\ Steinle-Neumann, L.\ %%@
Stixrude, V.\ Struzhkin for helpful discussions. S.\ Mukherjee performed technical assistance. This work was
supported by DOE ASCI/ASAP subcontract B341492 to Caltech DOE W-7405-ENG-48 and the National Science Foundation EAR-998002. %%@
Computations were performed on the Cray SV1 at the Geophysical Laboratory, supported by NSF grant EAR-9975753 and by the W.\ M.\ %%@
Keck Foundation.
\bibliographystyle{elsart-harv}
%\bibliography{pepihcp}

\begin{thebibliography}{27}
\expandafter\ifx\csname natexlab\endcsname\relax\def\natexlab#1{#1}\fi
\expandafter\ifx\csname url\endcsname\relax
  \def\url#1{\texttt{#1}}\fi
\expandafter\ifx\csname urlprefix\endcsname\relax\def\urlprefix{URL }\fi

\bibitem[{Asada and Terakura(1992)}]{asada}
Asada, T., Terakura, K., 1992. Cohesive properties of iron obtained by use of
  the generalized gradient approximation. Phys. Rev. B 46, 13599--13602.

\bibitem[{Cohen et~al.(1994)Cohen, Mehl, and Papaconstantopoulos}]{ronnonmag1}
Cohen, R., Mehl, M., Papaconstantopoulos, D., 1994. Tight-binding total energy
  method for transition and noble metals. Phys. Rev. B 50, 14694--14697.

\bibitem[{Cohen et~al.(1997)Cohen, Stixrude, and Wasserman}]{ronnonmag2}
Cohen, R., Stixrude, L., Wasserman, E., 1997. Tight-binding computations of
  elastic anisotropy of Fe, Xe, and Si under compression. Phys. Rev. B 56,
  8575--8589.

\bibitem[{Freyss et~al.(1997)Freyss, Stoeffler, and Dreysse}]{freyss}
Freyss, M., Stoeffler, D., Dreysse, H., 1997. Interfacial alloying and
  interfacial coupling in Cr/Fe(001). Phys. Rev. B 56, 6047.

\bibitem[{Jephcoat et~al.(1986)Jephcoat, Mao, and Bell}]{jephcoat}
Jephcoat, A.~P., Mao, H.-K., Bell, P.~M., 1986. Static compression of iron to
  78 GPa with rare-gas solids as pressure-transmitting media. J. Geophys. Res.
  91, 4677--4684.

\bibitem[{Krasko(1987)}]{krasko}
Krasko, G.~L., 1987. Metamagnetic behavior of fcc iron. Phys. Rev. B 36, 8565.

\bibitem[{Mao et~al.(1990)Mao, Wu, Chen, Shu, and Jephcoat}]{mao}
Mao, H.-K., Wu, Y., Chen, L., Shu, J., Jephcoat, A.~P., 1990. Static
  compression of iron to 300 GPa and Fe$_{0.8}$Ni$_{0.2}$ alloy to 260 GPa:
  Implications for composition of the core. J. Geophys. Res. 95, 21737--21742.

\bibitem[{Mehl et~al.(2001)Mehl, Papaconstantopoulos, Mazin, Bacalis, and
  Pickett}]{mehl}
Mehl, M.~J., Papaconstantopoulos, D., Mazin, I.~I., Bacalis, N., Pickett, W.,
  2001. Applications of the NRL tight-binding method to magnetic systems. J.
  Appl. Phys. 89, 6880--6882.

\bibitem[{Merkel et~al.(2000)Merkel, Goncharov, Mao, Gillet, and
  Hemley}]{merkel}
Merkel, S., Goncharov, A., Mao, H., Gillet, P., Hemley, R., 2000. Raman
  spectroscopy of iron to 152 gigapascals: Implications for Earth's inner core.
  Science 288, 1626--1629.

\bibitem[{Mryasov et~al.(1992)Mryasov, Gubanov, and Liechtenstein}]{fcc}
Mryasov, O., Gubanov, V., Liechtenstein, A., 1992. Spiral-spin-density-wave
  states in fcc iron: Linear-muffin-tin-orbitals band-structure approach. Phys.
  Rev. B 45, 12330--12336.

\bibitem[{Mukherjee and Cohen(2001)}]{sonali}
Mukherjee, S., Cohen, R.~E., 2001. Tight binding based non-collinear spin model
  and magnetic correlations in Iron. J. Comp.-Aid. Mat. Des. 8, 107--115.

\bibitem[{Perdew et~al.(1996)Perdew, Burke, and Ernzerhof}]{PBE}
Perdew, J., Burke, K., Ernzerhof, M., 1996. Generalized gradient approximation
  made simple. Phys. Rev. Lett. 77, 3865--3868, correction: ibid. 78, 1396
  (1997).

\bibitem[{Pickett(1996)}]{pickett}
Pickett, W., 1996. Non-collinear magnetic states: From density functional
  theory to model Hamiltonians. J. Korean Phys. Soc. 29, S70.

\bibitem[{Rueff et~al.(1999)Rueff, Krisch, Cai, Kaprolat, Hanfland, Lorenzen,
  Masciovecchio, Verbeni, and Sette}]{rueff}
Rueff, J.~P., Krisch, M., Cai, Y.~Q., Kaprolat, A., Hanfland, M., Lorenzen, M.,
  Masciovecchio, C., Verbeni, R., Sette, F., 1999. Magnetic and structural
  alpha-epsilon phase transition in Fe monitored by x-ray emission
  spectroscopy. Phys. Rev. B 60, 14510--14512.

\bibitem[{Sandratskii(1998)}]{sandratskii}
Sandratskii, M.~L., 1998. Noncollinear magnetism in itinerant-electron systems:
  theory and applications. Adv. Phys. 47, 91--160.

\bibitem[{Sj\"{o}stedt and Nordstr\"{o}m(2002)}]{sjostedt}
Sj\"{o}stedt, E., Nordstr\"{o}m, L., 2002. Noncollinear full-potential studies
  of $\epsilon$-Fe. Phys. Rev. B 66, 014447.

\bibitem[{Steinle-Neumann et~al.(1999)Steinle-Neumann, Stixrude, and
  Cohen}]{gerd}
Steinle-Neumann, G., Stixrude, L., Cohen, R., 1999. First-principles elastic
  constants for the hcp transition metals Fe, Co, and Re at high pressure.
  Phys. Rev. B 60, 791--799.

\bibitem[{Steinle-Neumann et~al.(2003)Steinle-Neumann, Stixrude, Cohen, and
  Kiefer}]{gerd2}
Steinle-Neumann, G., Stixrude, L., Cohen, R.~E., Kiefer, B., 2003. Evidence of
  local magnetic order in hcp iron from Raman mode splitting ,
  http://xxx.lanl.gov/abs/cond--mat/0111487.

\bibitem[{Stixrude et~al.(1994)Stixrude, Cohen, and Singh}]{bcc}
Stixrude, L., Cohen, R., Singh, D., 1994. Iron at high pressure: Linearized
  augmented plane wave computations in the generalized gradient approximation.
  Phys. Rev. B 50, 6442--6445.

\bibitem[{Stoner(1938)}]{stoner}
Stoner, E.~C., 1938. Collective electron ferromagnetism II. Energy and specific
  heat. Proc. R. Soc. London, Ser. A 169, 339--371.

\bibitem[{Taylor et~al.(1982)Taylor, Cort, and WIllis}]{taylor}
Taylor, R., Cort, G., WIllis, J., 1982. Internal magnetic fields in hcp-iron.
  J. Appl. Phys. 53, 8199--8201.

\bibitem[{Taylor et~al.(1991)Taylor, Pasternak, and Jeanloz}]{jeanloz}
Taylor, R., Pasternak, M., Jeanloz, R., 1991. Hysteresis in the high pressure
  transformation of bcc- to hcp-iron. J. Appl. Phys. 69, 6126--6128.

\bibitem[{Uhl et~al.(1994)Uhl, Sandratskii, and K\"{u}bler}]{uhl}
Uhl, M., Sandratskii, M.~L., K\"{u}bler, J., 1994. Spin fluctuations in
  $\epsilon$-Fe and in Fe$_{3}$ Pt Invar from local-density-functional
  calculation. Phys. Rev. B 50, 291--301.

\bibitem[{Ujfalussy et~al.(1999)Ujfalussy, Wang, Nicholson, Shelton, and
  Stocks}]{ujfalussy}
Ujfalussy, B., Wang, X.-D., Nicholson, D. M.~C., Shelton, W.~A., Stocks, G.~M.,
  1999. Constrained density functional theory for first principles spin
  dynamics. J. Appl. Phys. 85, 4824--4826.

\bibitem[{Vinet et~al.(1987)Vinet, Ferrante, Rose, and Smith}]{vinet}
Vinet, P., Ferrante, J., Rose, J., Smith, J., 1987. Compressibility of solids.
  J. Geophys. Res. 92, 9319--9325.

\bibitem[{Wasserman et~al.(1996)Wasserman, Stixrude, and Cohen}]{ronnonmag3}
Wasserman, E., Stixrude, L., Cohen, R., 1996. Thermal properties of iron at
  high pressures and temperatures. Phys. Rev. B 53, 8296--8309.

\bibitem[{You and Heine(1982)}]{you}
You, M.~V., Heine, V., 1982. Magnetism in transition metals at finite
  temperatures. I. Computational model. J. Phys. F 12, 177--94.

\end{thebibliography}

% \appendix

% \section{}
% \label{}

% Bibliographic references with the natbib package:
% Parenthetical: \citep{Bai92} produces (Bailyn 1992).
% Textual: \citet{Bai95} produces Bailyn et al. (1995).
% An affix and part of a reference:
%   \citep[e.g.][Ch. 2]{Bar76}
%   produces (e.g. Barnes et al. 1976, Ch. 2).

% \bibitem[Names(Year)]{label} or \bibitem[Names(Year)Long names]{label}.
% (\harvarditem{Name}{Year}{label} is also supported.)

\end{document}